# On the feasibility of pentamode mechanical metamaterials

Muamer Kadic[1], Tiemo Bückmann[1], Nicolas Stenger[1], Michael Thiel[2,3], and Martin Wegener[1,2]

[1] Institute of Applied Physics and DFG-Center for Functional Nanostructures (CFN), Karlsruhe Institute of Technology (KIT), 76131 Karlsruhe, Germany
[2] Institute of Nanotechnology (INT), Karlsruhe Institute of Technology (KIT), 76131 Karlsruhe, Germany
[3] Nanoscribe GmbH, Hermann-von-Helmholtz-Platz 1, 76344 Eggenstein-Leopoldshafen, Germany

Conceptually, all conceivable three-dimensional mechanical materials can be built from pentamode materials. Pentamodes also enable to implement three-dimensional transformation acoustics – the analogue of transformation optics. However, pentamodes have not been realized experimentally to the best of our knowledge. Here, we investigate inasmuch the pentamode theoretical ideal suggested by Milton and Cherkaev in 1995 can be approximated by a metamaterial with current state-of-the-art lithography. Using numerical calculations calibrated by our fabricated three-dimensional microstructures, we find that the figure of merit, i.e., the ratio of bulk modulus to shear modulus, can realistically be made as large as about 1,000.

Transformation optics can be seen as a design tool for steering light waves in a desired manner. In optics, one generally needs anisotropic magneto-dielectric (meta-) materials for, e.g., invisibility cloaks [1,2]. It is interesting to translate transformation optics to other types of waves such as acoustic waves. However, the three-dimensional elastodynamic equations are not invariant under coordinate transformations for scalar mass density and normal elastic materials [3]. In two dimensions or in thin plates, usual anisotropic elastic materials can suffice [4,5,6]. In three dimensions, one either needs materials with anisotropic mass density tensors [3,7,8,9] or pentamode materials [3,10,11,12] to implement the counterpart of invisibility cloaks or other devices. Neither of these materials has been realized experimentally so far.

In 1995, Milton and Cherkaev [13] showed that all conceivable mechanical materials can be synthesized on the basis of pentamodes. Pentamodes are special in the sense that they avoid the coupling of compression and shear waves by making the bulk modulus, $B$, extremely large compared to the shear modulus, $G$, ideally infinitely large [13,14]. This situation corresponds to isotropic fluids, for which $G = 0$ and thus the Poisson's ratio [15] is $\nu = \frac{3-2(G/B)}{2(G/B)+6} = 0.5$. Hence, pentamodes are sometimes also called "metafluids". Mathematically, 5 ("penta") of the 6 diagonal elements of the diagonalized $6 \times 6$ elasticity tensor of an isotropic pentamode material are zero, and only one is non-zero [13,14].

A conceptually perfect homogeneous pentamode material would literally immediately flow away. An intentionally spatially inhomogeneous pentamode structure would rapidly intermix and hence be destroyed, rendering these pentamode ideals essentially useless. Large metafluid viscosity could reduce these unwanted effects, but such internal friction would also introduce undesired damping/losses. Thus, in practice, one does want some *finite* shear modulus for stability. If the shear modulus is small compared to the bulk modulus, the ideas of transformation acoustics [3,7] are no longer exact, but are still expected to apply approximately. After all, perfect magneto-dielectrics have not been achieved in transformation optics either; nevertheless, striking results have been obtained with approximate materials [16,17].

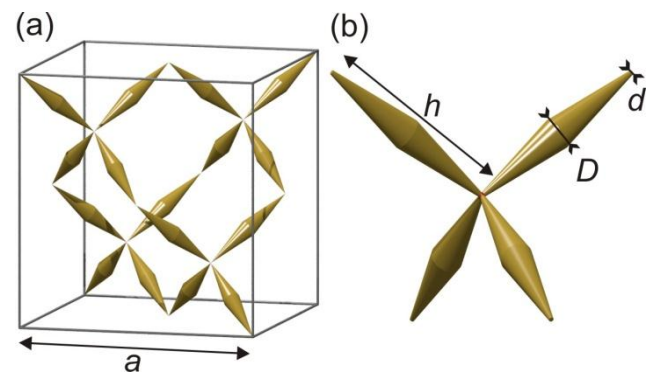

Fig.1: (a) Illustration of the pentamode metamaterial ideal suggested by Milton and Cherkaev in 1995. The artificial crystal with lattice constant $a$ has diamond symmetry. (b) Illustration of our approximation of the pentamode ideal by a stable three-dimensional microstructure. In contrast to (a), the connection regions of touching cones have a *finite* diameter $d$. The diameter of the thick end of the cones, $D$, and the total double-cone length, $h$, are also indicated.

For reference, regarding natural materials, bulk gold exhibits a fairly large ratio of bulk modulus to shear modulus of about $B/G = 13$, which will likely *not* be sufficient. In this letter, we show that $B/G$ ratios or "figures-of-merit (FOM)" in the range of $> 10^3$ are realistically accessible with metamaterial polymer microstructures made by dip-in direct laser writing [18].

Fig. 1(a) exhibits the pentamode metamaterial structure suggested by Milton and Cherkaev in 1995 [13]. It is composed of elements made by two connected truncated cones with total length $h$ (in vacuum). Ideally, these double-cone elements touch each other only at their strictly point-like tips, and form a diamond-type crystal, the mechanical metamaterial, with lattice constant

$$a = \frac{4h}{\sqrt{3}}.$$

It is known that the absolute size of the lattice constant does not matter in the static limit [15]. In contrast, for finite acoustic frequencies, the lattice constant needs to be small compared to the corresponding acoustic wavelength to qualify as an effective material. Thus, small lattice constants $a$ are desirable for high-frequency and broadband operation.

Unfortunately, however, it is obvious that with an infinitely small connection point the structure in Fig. 1(a) will fall apart upon only the slightest mechanical perturbation. In order to obtain stability, we consider a finite connection or overlap volume, replacing the strictly point-like tips above. This step is defined and illustrated in Figs. 1(b) and (c). It is clear that increasing the overlap volume will stabilize the structures, yet at the same time decrease the $B/G$ ratio. To study this trade-off quantitatively, we have performed numerical continuum-mechanics calculations using a commercial software package (COMSOL Multiphysics, MUMPS solver). The structure is composed of a constituent material, a typical polymer, in vacuum. The Young's modulus of the polymer has been taken as 3 GPa, its Poisson ratio as 0.4. However, for reasonable choices, the constituent material parameters will turn out to be not important for the pentamode $B/G$ ratio (see below). The pentamode bulk modulus, $B$, is directly obtained from its definition, i.e., from applying a hydrostatic pressure and observing the resulting compression within the linear regime (i.e., strains smaller than 1%). Precisely, "hydrostatic pressure" means that we push on all polymer ends at all six interfaces of the metamaterial cube in free space normal to the respective interface. The shear modulus, $G$, is obtained by applying a shear force and calculating the displacement. The calculations have been obtained for pentamode structures composed of $N \times N \times N$ pentamode unit cells like shown in Fig. 1(a) with up to $N = 5$ or 125 unit cells total, discretized by about $10^6$ tetrahedra in three dimensions on an adaptive mesh. Interestingly, the results are nearly identical even for a single unit cell, i.e., for $N = 1$. This provides us with confidence that we actually compute "bulk" metamaterial properties.

The resulting dependence of the bulk and shear modulus as well as the $B/G$ ratio on the small diameter, $d$, are depicted in Fig. 2. Selected connection regions are shown as insets in (b) for illustration. As expected from our above reasoning, the $B/G$ ratio increases with decreasing diameter (see Fig. 2(b)). It actually diverges for $d \to 0$. For clarity, we have thus chosen a double-logarithmic representation in Fig. 2. At small values of $d$, the $B/G$ ratio scales approximately like $B/G \propto 1/d^2$ (see corresponding global fit in Fig. 2(b)). The increase in $B/G$ with decreasing $d$ mainly stems from a more rapid decrease of $G$ compared to that of $B$ (see Fig. 2(a)).

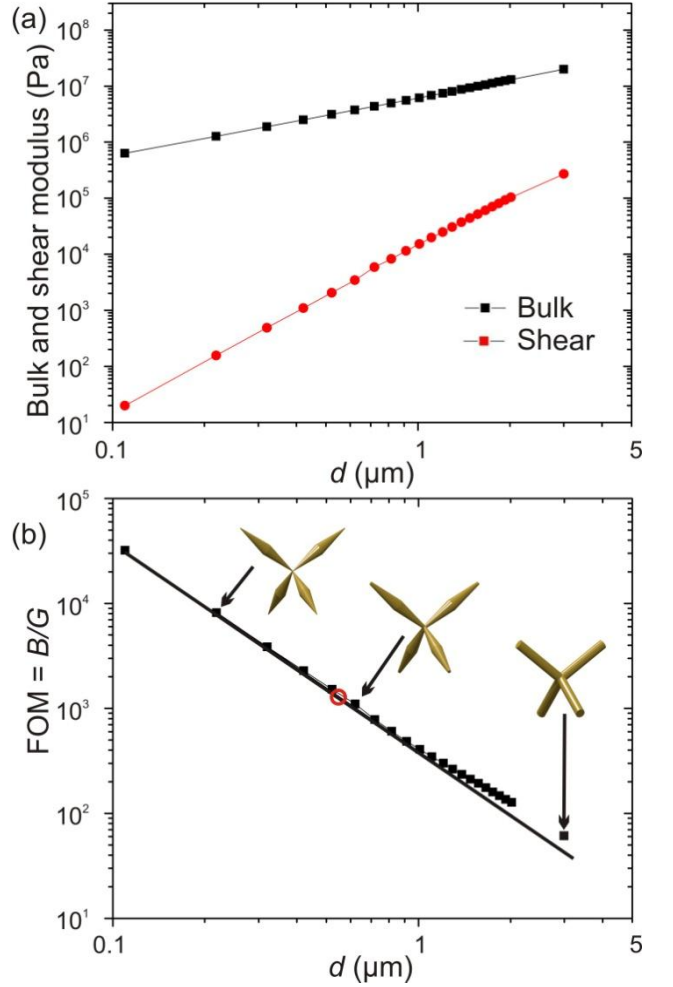

Fig.2: (a) Bulk modulus, $B$, and shear modulus, $G$, versus the diameter, $d$, of the thin end of the cones (see Fig. 1) in a double-logarithmic representation. (b) Resulting figure of merit (FOM), i.e., the ratio $B/G$. The other parameters are fixed to $D = 3$ μm and $h = 16.15$ μm. The red circle at $d = 0.55$ μm corresponds to the experiment shown in Fig. 3. At this point, we approximately obtain $B/G = 10^3$; at $d = 0.1$ μm, we even find values exceeding $B/G = 10^4$. The insets illustrate the corresponding pentamode metamaterial structures at selected values of $d$ (see arrows).

To calibrate our calculations in regard to feasible structures and to realistically experimentally accessible smallest connection diameters $d$, Figs. 3 and 4 show polymer structures which we have fabricated using dip-in direct-laser-writing (DLW) optical lithography. The application of this new technology to other three-dimensional mechanical metamaterials has recently been described in detail elsewhere [18] and shall not be repeated here. In brief, in contrast to regular DLW [19], we use the photoresist as the immersion fluid directly on the microscope objective lens [18]. The fabricated structures may not yet be fully optimized, but they are stable and reveal high sample quality. Fig. 3 exhibits an small diameter of only about $d = 0.55$ μm. This value is indicated by the red circle in Fig. 2(b). At this point, the calculated $B/G$ ratio is already larger than $10^3$. For more aggressive values of $d = 0.1$ μm, the $B/G$ ratio exceeds $10^4$. We note in passing that the values for $B$ and hence for $B/G$ obtained for fixing the metamaterial sample cube to a substrate, by additionally fixing the four side walls, and for only pushing onto the top surface differ by less than 10%. This finding is relevant in that future experiments might be easier to conduct along these lines. Fig. 4 highlights that truly three-dimensional structures with appropriate symmetry can be achieved experimentally.

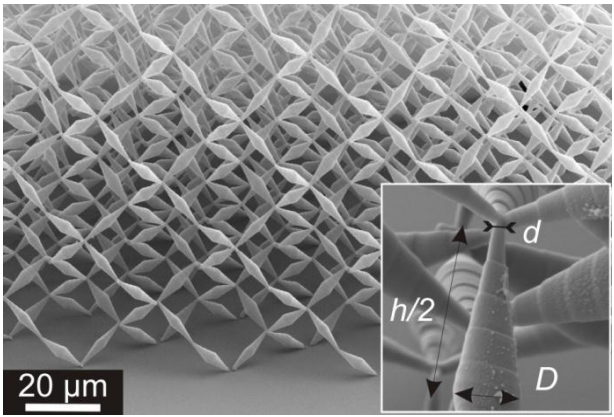

Fig.3: Oblique-view electron micrograph of a polymer pentamode mechanical metamaterial fabricated by dip-in three-dimensional direct-laser-writing (DLW) optical lithography on a glass substrate with $h = 16.15$ μm and $D = 3$ μm (compare Fig. 1(b)). Importantly, the inset shows that the smallest accessible diameter of the connection regions of adjacent cones is about $d = 0.55$ μm. This value is highlighted by the red circle in Fig. 2(b).

Figure 5(a) shows the calculated dependence of $B$ and $G$ on the large diameter $D > d$ for three different fixed values of $d$. Both moduli change because $D$ directly influences the angle of the cones and hence the shape of their thin ends, which dominate the mechanical properties. The resulting $B/G$ ratio shown in Fig. 5(b) exhibits a fairly weak dependence on the large diameter, $D$, which can be fitted by $B/G \propto 1/\sqrt{D}$ (see corresponding global fit).

We have also studied the dependence of the $B/G$ ratio on the properties of the constituent material. Upon decreasing the Poisson's ratio of the constituent material from 0.4 (used so far) to 0.1 while fixing the Young's modulus, the pentamode $B/G$ ratio typically increases by 20% (not depicted). As a very different example we have also calculated fictitious pentamodes made of diamond (Young's modulus taken as 1220 GPa and Poisson's ratio taken as 0.2) instead of polymer. The resulting $B/G$ ratios (not depicted) differ from the polymer case by not more than 10%. We conclude that the properties of the constituent material (if stable itself) have little effect on the resulting metamaterial $B/G$.

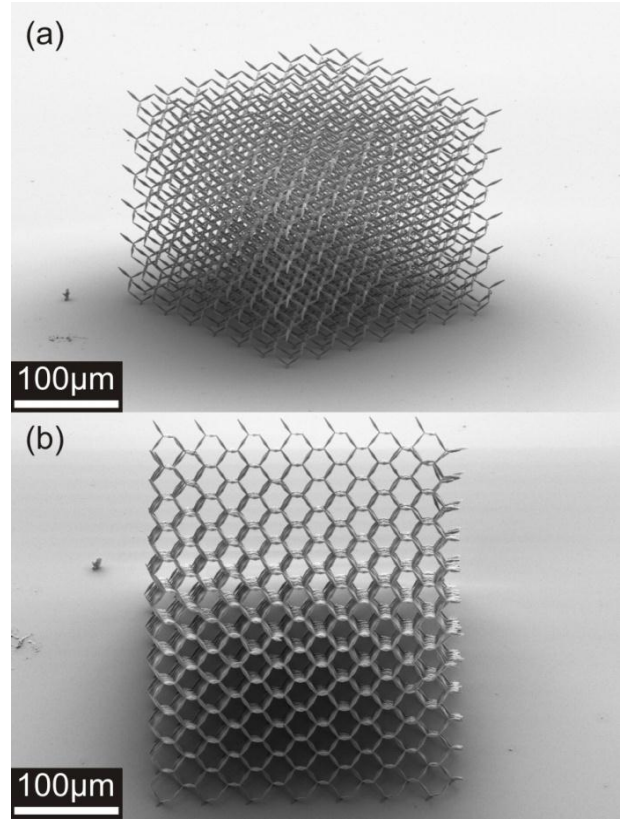

Fig.4: (a) Oblique-view electron micrograph of another polymer pentamode mechanical metamaterial with $7 \times 7 \times 6$ face-centered-cubic unit cells (hence total size 261 μm × 261 μm × 224 μm) and with $h = 16.15$ μm, $D = 3$ μm, and $d = 1$ μm fabricated by dip-in DLW. (b) Same structure, but view along the (001)-direction to reveal the hexagonal substructure of the diamond crystal. We estimate a volume filling fraction of 1.5% and a total metamaterial weight of 270 ng.

Heuristically, we can summarize all of our numerical findings for pentamode structures like shown in Fig. 1(b) (i.e., for $d < D \ll h$) within the linear regime by the simple approximate universal formula

$$\text{FOM} = \frac{B}{G} = \left(\frac{h}{d}\right)^2 \sqrt{\frac{h}{D}}.$$

We have used this formula for all "global fits" shown by the solid curves in Fig. 2(b) and Fig. 5(b). In the ideal pentamode limit of $B/G \gg 1$, it is straightforward to derive the resulting asymptotic form of the Poisson's ratio $\nu = \frac{1}{2}\left(1 - \left(\frac{B}{G}\right)^{-1}\right)$. As expected [15], these expressions do not depend on any absolute scale, i.e., if all structure dimensions ($d$, $D$, and $h$) are multiplied by the same factor, the $B/G$ ratio and $\nu$ remain unaltered. However, as pointed out above, this static result is only expected to be valid for finite frequencies as long as the corresponding acoustic wavelength is large compared to the pentamode metamaterial lattice constant $a$. Quantitatively identifying this maximum frequency, which depends on the constituent material, requires phonon band structure calculations of the pentamode metamaterial, which have not been published to the best of our knowledge. Their computation is interesting and relevant, but far beyond the scope of the present paper.

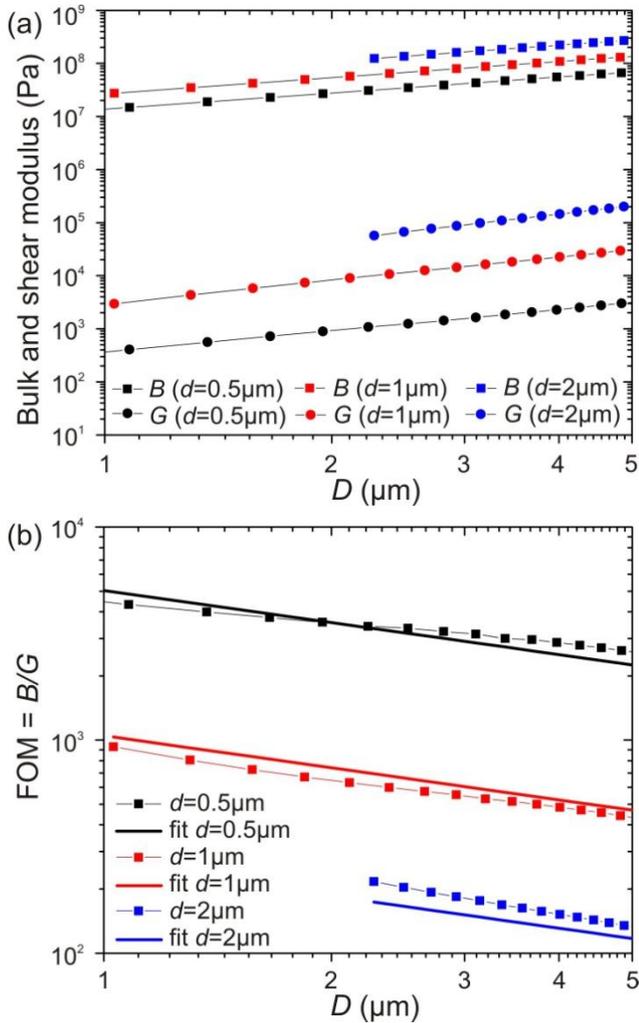

Fig.5: (a) Bulk modulus $B$ and shear modulus $G$ as a function of the diameter of the thick end of the cones $D$ (see Fig. 1(b)), for fixed $h = 16.15$ μm and for three different fixed values of $d$ as indicated. All other parameters and the representation are as in Fig. 3. (b) Resulting FOM = $B/G$. The solid straight lines are global fits according to $B/G \propto 1/\sqrt{D}$

In conclusion, we have shown that state-of-the-art three-dimensional laser lithography enabling critical dimensions of $d = 0.55$ μm should allow for pentamode three-dimensional mechanical metamaterials with double-cone lengths in the ten-micrometer range and with figures of merit beyond $10^3$. This step potentially enables three-dimensional transformation-acoustics architectures. While we have successfully demonstrated the experimental making of such pentamode structures, their direct and detailed experimental characterization and application to three-dimensional transformation acoustics poses demanding, yet also very appealing future challenges, too. If the critical dimensions could be reduced from 0.55 μm to 0.1 μm while maintaining the other parameters, even pentamode figures of merit exceeding $10^4$ might become possible in the future.

We thank Johannes Kaschke for help with the electron micrographs. We acknowledge support by the Deutsche Forschungsgemeinschaft (DFG), the State of Baden-Württemberg, and the Karlsruhe Institute of Technology (KIT) through the DFG-Center for Functional Nanostructures (CFN) within subprojects A1.4 and A1.5. N.S. received support by the Karlsruhe School of Optics & Photonics (KSOP).

[1] J. B. Pendry, D. Schurig, and D. R. Smith, Science **312**, 1780 (2006).
[2] U. Leonhardt and T. G. Philbin, *Geometry and Light: The Science of Invisibility* (Dover, Mineola, 2010).
[3] G. W. Milton, M. Briane, and J. R. Willis, New J. Phys. **8**, 248 (2006).
[4] M. Farhat, S. Guenneau, and S. Enoch, Phys. Rev. Lett. **103**, 024301 (2009).
[5] N.V. Movchan, R. C. McPhedran, and A. B. Movchan, Proc. R. Soc. A **467**, 869 (2010).
[6] N. Stenger, M. Wilhelm, and M. Wegener, Phys. Rev. Lett. **108**, 014301 (2012).
[7] A. N. Norris, Proc. R. Soc. A **464**, 2411 (2008).
[8] H. Chen and C. T. Chan, Appl. Phys. Lett. **91**, 183518 (2007).
[9] S. A. Cummer, B.-I. Popa, D. Schurig, D. R. Smith, J. Pendry, M. Rahm, and A. Starr, Phys. Rev. Lett. **100**, 024301 (2008).
[10] A. N. Norris, J. Acoust. Soc. Am. **125**, 839 (2009).
[11] P. H. Mott, J. R. Dorgan, and C. M. Roland, J. Sound Vibr. **312**, 572 (2008).
[12] C. L. Scandrett, J. E. Boisvert, and T. R. Howarth, J. Acoust. Soc. Am. **127**, 2856 (2010).
[13] G. W. Milton and A. Cherkaev, J. Eng. Mater. Technol. **117**, 483 (1995).
[14] G. W. Milton, *The Theory of Composites* (Cambridge: Cambridge University Press, 2002).
[15] G. N. Greaves, A. L. Greer, R. S. Lakes, and T. Rouxel, Nature Mater. **10**, 823 (2011).
[16] D. Schurig, J. J. Mock, B. J. Justice, S. A. Cummer, J. B. Pendry, A. F. Starr, and D. R. Smith, Science **314**, 977 (2006).
[17] T. Ergin, N. Stenger, P. Brenner, J.B. Pendry, and M. Wegener, Science **328**, 337 (2010).
[18] T. Bückmann, N. Stenger, M. Kadic, J. Kaschke, A. Frölich, T. Kennerknecht, C. Eberl, M. Thiel, and M. Wegener, submitted (2012).
[19] G. von Freymann, A. Ledermann, M. Thiel, I. Staude, S. Essig, K. Busch, and M. Wegener, Adv. Funct. Mater. **20**, 1038 (2010).